\documentclass[
twocolumn,
superscriptaddress,
aps,
prl,
]{revtex4-2}

\usepackage[T1]{fontenc}
\usepackage[utf8]{inputenc}
\usepackage[english]{babel}
\addto\captionsenglish{}
\addto\captionsenglish{}
\usepackage{graphicx}
\usepackage[colorlinks,breaklinks]{hyperref}
\usepackage{mathtools}
\usepackage{bm}
\usepackage{amssymb}
\usepackage{overpic}
\usepackage[section]{placeins}
\usepackage{enumitem}
\usepackage{microtype}
\usepackage{xcolor}

\makeatletter
\g@addto@macro\@floatboxreset{\centering}
\makeatother

\graphicspath{{figures/}}

\renewcommand{\refeq}[1]{Eq.~(\ref{eq:#1})}
\newcommand{\reffig}[1]{Fig.~\ref{fig:#1}}
\newcommand{\reffigs}[2]{Figs.~\ref{fig:#1} and~\ref{fig:#2}}
\newcommand{\reftab}[1]{Table~\ref{tab:#1}}

\newcommand{\refref}[1]{Ref.~\cite{#1}}

\newcommand{\Rey}{\ensuremath{\mathrm{Re}}}
\newcommand{\Reystar}[2]{\ensuremath{\Rey^\mathrm{#1}_\mathrm{#2}}}
\newcommand{\Reycr}{\ensuremath{\Rey_\mathrm{cr}}}

\newcommand{\Reypatch}{\ensuremath{\Reystar{full}{STI}}}
\newcommand{\Reytransient}{\ensuremath{\Reystar{STI}{trans}}}
\newcommand{\Reyattractor}{\ensuremath{\Reystar{trans}{chaos}}}
\newcommand{\Reytorus}{\ensuremath{\Reystar{chaos}{torus}}}
\newcommand{\ReysRPO}{\ensuremath{\Reystar{torus}{PO}}}
\newcommand{\ReyuRPO}{\ensuremath{\Reystar{PO}{uPO}}}
\newcommand{\ReyTW}{\ensuremath{\Reystar{uPO}{TW}}}
\newcommand{\Reysaddlenode}{\ensuremath{\Reystar{TW}{lam}}}

\begin{document}

\title{Direct path from turbulence to time-periodic solutions}

\author{Chaitanya S. Paranjape}
\thanks{C.P.\ and G.Y.\ contributed equally to this work.}
\affiliation{Institute of Science and Technology Austria (ISTA), 3400 Klosterneuburg, Austria}
\author{Gökhan Yalnız}
\thanks{C.P.\ and G.Y.\ contributed equally to this work.}
\affiliation{Institute of Science and Technology Austria (ISTA), 3400 Klosterneuburg, Austria}
\author{Yohann Duguet}
\affiliation{LISN-CNRS, Campus Universitaire d'Orsay, Universit\'e Paris-Saclay, 91405 Orsay, France}
\author{Nazmi Burak Budanur}
\affiliation{Institute of Science and Technology Austria (ISTA), 3400 Klosterneuburg, Austria}
\affiliation{Max Planck Institute for the Physics of Complex Systems (MPIPKS), 01187 Dresden, Germany}
\author{Björn Hof}
\affiliation{Institute of Science and Technology Austria (ISTA), 3400 Klosterneuburg, Austria}

\date{June 7, 2023}

\begin{abstract}
Viscous flows through pipes and channels are steady and ordered until, with increasing velocity, the laminar motion catastrophically breaks down and gives way to turbulence. How this apparently discontinuous change from low- to high-dimensional motion can be rationalized within the framework of the Navier--Stokes equations is not well understood. Exploiting geometrical properties of transitional channel flow we trace turbulence to far lower Reynolds numbers (\Rey) than previously possible and identify the complete path that reversibly links fully turbulent motion to an invariant solution. This precursor of turbulence destabilizes rapidly with \Rey, and the accompanying explosive increase in attractor dimension effectively marks the transition between deterministic and \emph{de facto} stochastic dynamics.
\end{abstract}
\maketitle

The origin of turbulence in pipe and channel flows has been debated for over a century. In recent years much effort has been dedicated to link the formation of turbulence to simple invariant solutions of the governing Navier--Stokes equations (periodic orbits, equilibria and traveling waves), which are commonly referred to as exact coherent structures (ECS) \cite{kerswell2005recent,eckhardt2007turbulence}. ECSs are suggested as building blocks of the turbulent dynamics \cite{kawahara2012significance,budanur2017relative,suri2020capturing,yalniz2021coarse,hof2004experimental}. However, efforts to directly link specific ECSs to the turbulent state, let alone to identify a reversible path connecting the two, have remained so far unsuccessful. While specific ECSs have been identified as starting points of bifurcation sequences into chaos \cite{kreilos2012periodic,avila2013streamwise,zammert2015crisis,ritter2016emergence,lustro2019onset}, the traceable path in parameter space towards turbulence in all these cases ends at a \emph{boundary crisis} \cite{grebogi1982chaotic}. At this point the attractor ceases to exist, giving way to short-lived transient chaos. Although a sufficiently fast ramp up in \Rey{} will prevent relaminarization and lead to turbulence, strictly, this only shows that the chosen route leads to the basin of attraction of turbulence \cite{duguet2008transition}. It does not necessarily prove, however, that the turbulent state originates from the specific ECSs, e.g.\ via a sequence of bifurcations. An unambiguous way to determine its roots would require starting directly from the turbulent state and tracing it quasi-statically down to its origin, a path prohibited by the aforementioned relaminarization barrier. 

This situation is markedly different from simpler transition scenarios encountered e.g.\ in supercritical Taylor--Couette flow and Rayleigh--B\'enard convection. In such cases a linear instability of the base flow gives rise to a \emph{primary vortex state}, which is the starting point of the bifurcation sequence leading to chaotic and eventually high-dimensional, turbulent motion. In particular, signatures of the primary vortex state tend to persist and can be detected in turbulent flow fields at values of \Rey{} several orders of magnitude larger than the instability threshold \cite{lathrop1992transition}. Hence, the role of the primary state and the connection with the subsequent dynamics is without question.

The purpose of the present study is to unambiguously identify the equivalent of the primary vortex state in aforementioned linearly stable flows, i.e.\ to determine the precursor turbulence originates from for channel flow. While given the transient nature of turbulence this may appear unfeasible, we show that by bypassing the regime of fully localized turbulent structures, turbulence can be traced beyond the transient regime all the way to its origin. The reverse path towards fully turbulent flow extends across a considerable \Rey{} range. However, surprisingly, stochasticity arises directly at the outset of this route, when the dimension increases explosively across a minute variation in parameter. 

\begin{figure}
    \includegraphics[width=0.99\linewidth]{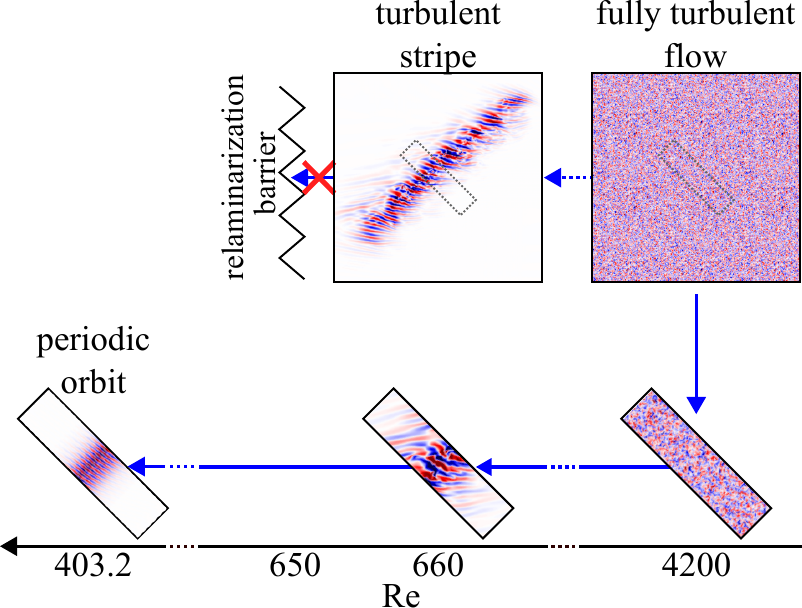}
    \caption{Sketch of the \Rey{} descent in a tilted domain vs.\ large non-tilted domains.
    Shown are wall-normal velocity ($v$) contours (range limited from $v=-0.1$ in blue to $v=0.1$ in red) in the midplane for various values of \Rey{}. Flow is from left to right. Dashed rectangles indicate the size and orientation of the tilted domain.
    (Bottom row images) $10\times40$ sized tilted domain investigated in this work.
    (Top row images) $100\times100$ crops from non-tilted domains (display scale is half that of bottom row images).}
    \label{fig:cartoon}
\end{figure}

Turbulence is space-filling at sufficiently large \Rey{} (\Rey{} is based here on the half-gap $h$, the kinematic viscosity and the laminar centerline velocity). At lower velocities turbulence becomes spatio-temporally intermittent (STI) and tends to organize in stripes interspersed with laminar regions \cite{shimizu2019bifurcations,paranjape2019thesis,kashyap2020flow,kashyap2022linear}. Below $\Rey{} \approx 650$ \cite{mukund2021aging} stripes are short-lived. Under standard circumstances this transient nature prevents continuation of turbulence towards lower \Rey{} (see \reffig{cartoon} top row) and prohibits further insights into its dynamical origin. 

In an attempt to circumvent this problem, we carry out direct numerical simulations in a domain that, on the one hand, is sufficiently large to capture generic turbulence at high \Rey{} and, on the other hand, is of the minimal size to capture turbulent stripes of a prescribed angle. Such minimal flow units for stripes \cite{tuckerman2020patterns} make use of the freedom to choose the orientation of the computational domain in the periodic directions. In our case we selected a tilt angle of 45 degrees with respect to the streamwise direction. Owing to the periodic boundary conditions stripes align at this prescribed angle (see \reffig{cartoon} bottom row for examples). This choice of a 45 degree angle is motivated by channel experiments where the same orientation is observed for stripes close to the onset of turbulence \cite{paranjape2019thesis}. The selected domain size, as in \cite{tuckerman2014turbulent}, is $10 \times 40$ ($L_x \times L_z$) in units of $h$. The incompressible Navier--Stokes equations are advanced in time using a standard spectral method in a three-dimensional domain with periodic boundary conditions in the plane and no-slip at the walls, with constant mass flux \cite{willis2017openpipeflow, xiao2020growth, channelflow2}.

The simulations started from a fully turbulent flow field at $\Rey=4200$ (corresponding to a friction Reynolds number of $Re_{\tau}=180$ \cite{kim1987turbulence}). \Rey{} was subsequently reduced in several steps down to $ \Reypatch{} \coloneqq 1500$, where turbulence becomes patterned \footnote{See Supplemental Material (SM) at [URL will be inserted by publisher] for (i) a movie illustrating the descent process, (ii) relation of the tilted to non-tilted coordinate systems, (iii) further details on invariant solutions, (iv) bifurcation sequence in a larger domain, and (v) high \Rey{} simulations in tilted and non-tilted domains.} (see \reftab{reynolds} for a list of transitions encountered for decreasing \Rey{}). From here the descent was continued in small steps, allowing the flow to settle for 500 advective time units (ratio of $h$ by the laminar centerline velocity) between consecutive steps. Below $\Rey=900$ the step size was set to $\Delta \Rey=2$. Typical adjustments of the turbulent flow occur within less than 100 advective time units. 
Once \Rey{} falls below $\Reytransient{}\approx 650$, and in agreement with the aforementioned experimental observations\cite{mukund2021aging} stripes are found to decay. However, in the present case lifetimes remain much larger and typically exceed several thousand advective time units. In agreement with these recent experiments we hence propose that $\Reytransient{}\approx 650$ is close to the point above which turbulence in extended domains (large $L_z$) first becomes sustained. In our tilted domain simulations lifetimes below this threshold remain sufficiently long for turbulence to reach what can be considered a statistically quasi-steady state.
\begin{figure}
\begin{overpic}[width=0.99\linewidth]{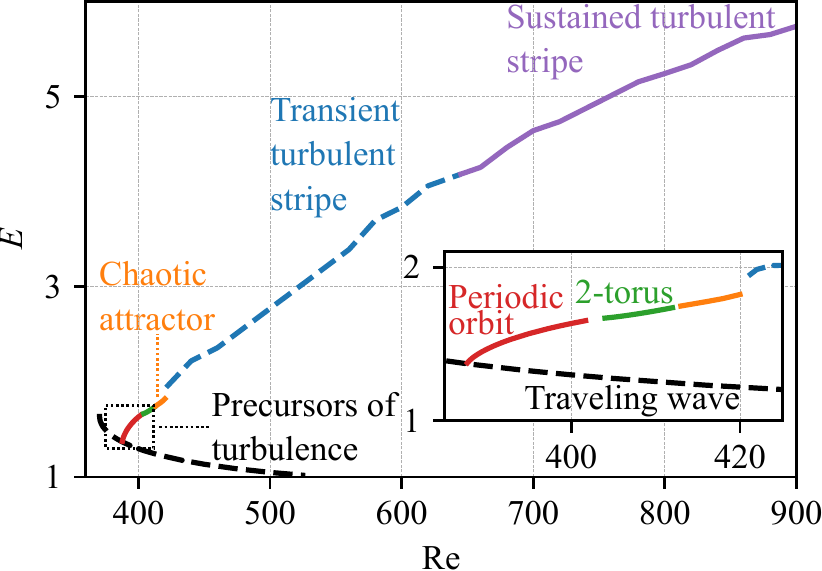}
    \put (0,70) {(a)}
\end{overpic}\\
\vspace{4mm}
\begin{overpic}[width=0.99\linewidth]{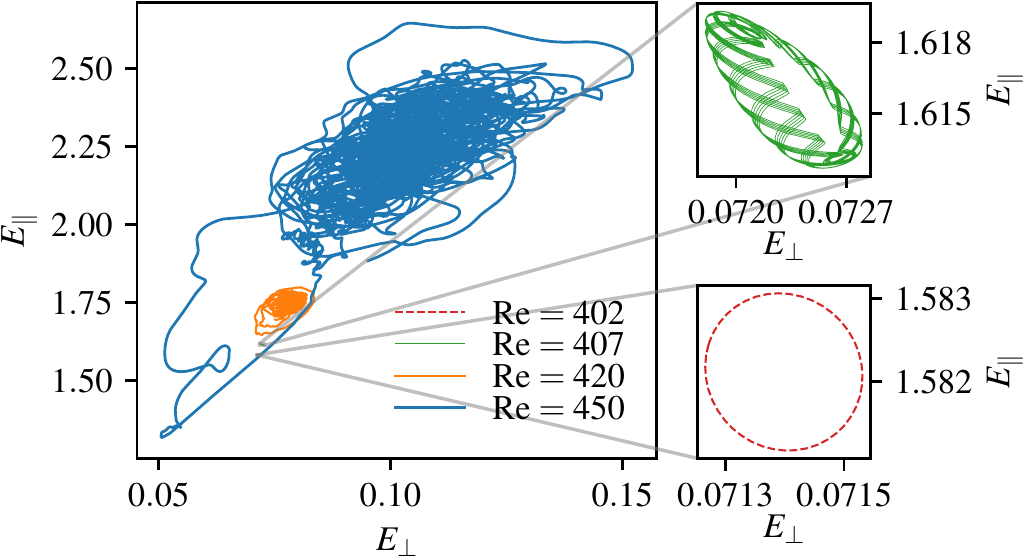}
    \put (0,55.4) {(b)}
    \put (65,55.4) {(c)}
    \put (65,28) {(d)}
\end{overpic}
\caption{Stripe turbulence during the \Rey{} descent. (a) Time-averaged perturbation kinetic energy $E$ vs.\ \Rey{}. Annotations refer to the respective dynamics encountered as \Rey{} is decreased. (b) $(E_\perp, E_\parallel)$ phase portraits of the instantaneous dynamics at various \Rey{} demonstrating some of the dynamics annotated in (a), with zoomed-in panels for (c) the torus at $\Rey=407$ and (d) the periodic orbit at $\Rey=402$. $E_\perp$ is the kinetic energy associated with the wall-normal component ($v$) of perturbation velocity $\bm{u}$, $E_\perp=\int_{V} v^2/2\, \mathrm{d}V$, and $E$ the perturbation kinetic energy, $E=\int_{V} \bm{u} \cdot \bm{u}/2\, \mathrm{d}V$, where $V$ is the computational domain. $E_\parallel=E-E_\perp$ is the kinetic energy associated with the in-plane components of perturbation velocity.}
\label{fig:bifurcations}
\end{figure}

As shown in the Supplemental Movie \cite{Note1}, the thereby stabilized stripe is followed far below \Reytransient{}, a regime previously inaccessible in experiments and simulations. With decreasing \Rey{} the perturbation kinetic energy of the stripe reduces (see \reffig{bifurcations}(a)), nevertheless fluctuations remain large and the flow is strongly chaotic even for \Rey{} as low as 450. For lower \Rey{}, as attested by the phase portrait of the dynamics in \reffig{bifurcations}(b), fluctuations reduce fast in amplitude and the state space region explored by the chaotic dynamics shrinks substantially. Eventually, the dynamics ceases to be chaotic (\Reytorus{}) and instead becomes quasiperiodic: the trajectory evolves on a 2-torus in state space (\reffig{bifurcations}(c)), and below \ReysRPO{} becomes periodic (see \reffig{bifurcations}(d) and the Supplemental Movie \cite{Note1}). The previously turbulent stripe hence simplifies to an exact coherent structure (\reffig{cartoon}). Despite its dynamical simplicity, the key spatial features, such as streamwise localization, characteristic spacing of streaks and vortices, associated large-scale flow parallel to the interface \cite{duguet2013oblique}, have been preserved all along this reduction in \Rey. As we further discuss in the SM, the periodic orbit (PO) can be continued \cite{viswanath2007recurrent,channelflow2} to even lower \Rey{}. It is shown to originate from a lower branch traveling wave, an edge state previously identified in \cite{paranjape2019thesis,paranjape2020oblique}. To probe the robustness of this transition scenario, we repeated the descent in a much larger ($10\times120$) domain and observed the same bifurcation sequence. A notable aspect of the above \Rey{} reduction is the sudden decrease of the attractor size at the final stages of the approach to the PO: the energy fluctuations displayed by the turbulent stripe at $\Rey=450$ (\reffig{bifurcations}(b)) are more than two orders of magnitude larger than those of the PO at $\Rey=402$ (\reffig{bifurcations}(d)). 
\begin{table}
	\caption{List of Reynolds numbers, at which the dynamics changes qualitatively. The second column specifies the dynamics observed below the given \Rey{} in simulations
    in the $10 \times 40$ sized tilted domain.
    Subscripts refer to the observations below the given \Rey{} whereas superscripts refer to the observations above.}
    \label{tab:reynolds}
	\begin{ruledtabular}
		\begin{tabular}{l r l}
            Name & Value & Observations below \Rey{} \\
            \hline
            \Reypatch{}      & $\approx 1500$   & Spatio-temporal intermittency\\[1mm]
            \Reytransient{}  & $\approx 650$    & Transient chaos\\[1mm]
            \Reyattractor{}  & $\approx 420$    & Sustained chaos\\[1mm]
            \Reytorus{}      & $412.8$          & Quasiperiodicity\\[1mm]
            \ReysRPO{}       & $403.2$          & Periodic orbits\footnote{See SM \cite{Note1} for a table continued beyond this point.}
		\end{tabular}
	\end{ruledtabular}
\end{table}

\begin{figure}
\includegraphics[width=0.99\linewidth]{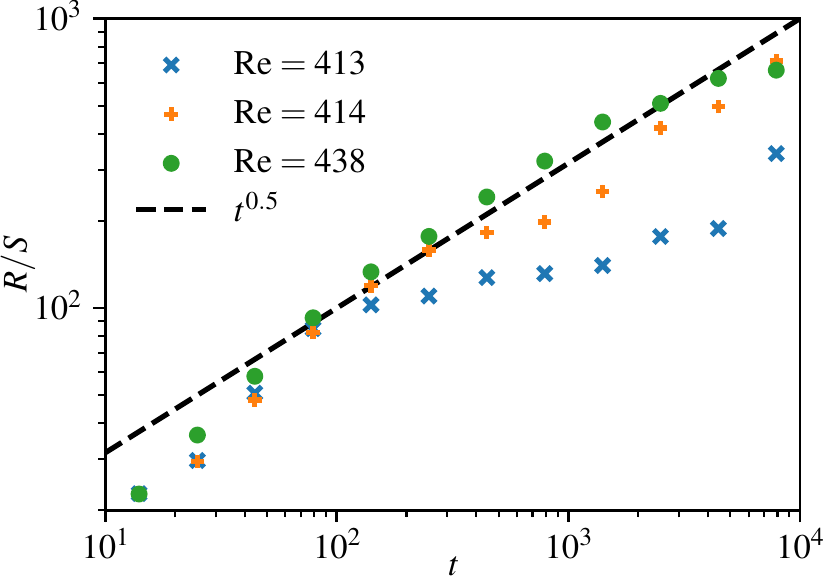}
\caption{$R/S$ vs.\ time in log-log scale, the slope of which defines the Hurst exponent (H).
For $\Rey=438$, a slope of 1/2 is approached after 100 advective time units. $H=1/2$ is expected for stochastic time series.}
\label{fig:Hurst}
\end{figure}

To obtain a better understanding of the emergence of turbulence, we take the PO as the starting point and investigate how the dynamics unfolds in the reverse direction, i.e.\ with increasing \Rey. To this end we analyze time series of the perturbation kinetic energy $E$. The method assumes the knowledge of consecutive values $E_0, E_1, \dots, E_{n-1}$, sampled every $\Delta t$. We first monitor the Hurst exponent $H(\Rey)$ associated with this time series of length $n$. $H$ quantifies the correlation of a signal, and is defined as the exponent in the scaling relation
\begin{equation}
\mathbb{E}(R/S)\sim n^H,\, n \rightarrow \infty\,,
\label{eq:hurst}
\end{equation}
where $R$ is the range of the first $n$ cumulative deviations from the mean, $S$ the sum of the first $n$ standard deviations, and $\mathbb{E}$ stands for the expected value \cite{hurst1951long}. The quantity $D=2-H$ is interpreted as a fractal dimension, namely that of the related signal \cite{mandelbrot1982fractal}. As shown in \reffig{Hurst}, the stochastic limit of $H = 1/2$ is already observed for a time horizon of 100 advective units at $\Rey=438$, i.e.\ in the transient regime far below the onset of sustained turbulence \cite{mukund2021aging}. This testifies that, for time horizons larger than 100 advective time units, the time series is indistinguishable from a purely stochastic signal. $O(100)$ advective time units match the typical time for localized turbulent structures to lose their memory after being created from a disturbance \cite{hof2006finite} (this timescale is also referred to as $t_0$ in lifetimes studies \cite{hof2008repeller}).

\begin{figure}
\includegraphics[width=0.99\linewidth]{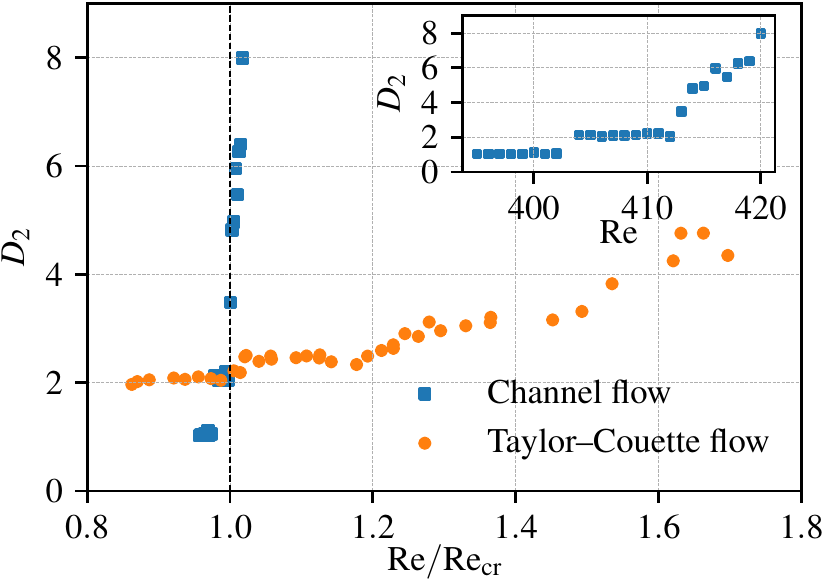}
\caption{Correlation dimension $D_2$ vs.\ \Rey{}. Blue squares are computed from time series of plane channel flow. Orange circles are from the turbulent Taylor--Couette flow experiments in \refref{brandstater1987strange}. \Reycr{} denotes the onset of chaos.}
\label{fig:D2}
\end{figure}

As a further estimation of the trend towards stochasticity, we compute another fractal dimension, the correlation dimension $D_2$ of the full turbulent set, using the Grassberger--Procaccia algorithm \cite{grassberger1983characterization,hegger1999practical}. For any integer $m>0$ and any real $\varepsilon>0$, $C_m(\varepsilon)$ is defined by
\begin{equation}
C_m(\varepsilon)=\lim_{n \rightarrow \infty}\frac{1}{n^2}\sum_{(j,k)}^{n}\Theta(\varepsilon - ||{\bm s}_{j}-{\bm s}_{k}||_m)\,,
\label{eq:C}
\end{equation}
where $\bm{s}_k=(E_{k-(m-1)\tau},\dots,E_{k-\tau},E_{k})$ is a delay vector in the $m$-dimensional embedded space, $||\cdot||_m$ a norm in that space, and $\Theta$ the Heaviside function. $\tau>0$ is a finite time delay expressed in \refeq{C} in units of the sampling time $\Delta t$ (in practice $\tau=60$ advective time units, close to the correlation time). $C_m$ counts temporal near-recurrences in the $m$-dimensional embedded space. The dimension $D_2$ is fitted as the exponent, for large $m$, in the scaling relation
\begin{equation}
C_m(\varepsilon) \sim \varepsilon^{D_2},\, \varepsilon \rightarrow 0\,.
\label{eq:D2}
\end{equation}
The amount of uncorrelated data necessary for the estimation of $D_2$ rises exponentially with its value \cite{eckmann1992fundamental}, which in practice limits computations to values below 10. $D_2$ is computed here starting from the PO at $\Rey=395$ up to $\Reyattractor=420$ deeper into the chaotic regime (blue squares in \reffig{D2}).

\reffig{D2} shows $D_2$ as a function of $\Rey/\Reycr$, where \Reycr{} stands for the onset of chaotic dynamics. For channel flow, \Reycr{} is identical to $\Reytorus=412.8$. Temporal chaos develops at a rapid pace: exceeding \Reycr{} by just over 1\% causes the dimension to quadruple. This rapid increase is consistent with the high dimension estimated for turbulent channel flows for larger \Rey{} \cite{keefe1992dimension}. For comparison we show the classical case of the supercritical transition in Taylor--Couette flow \cite{brandstater1987strange} (orange circles in \reffig{D2}) where even a 50\% increase above \Reycr{} results in only a doubling of the dimension.

The explosive dimension increase encountered in channel flow sharply limits the forecasting horizon directly at the onset of chaos, and decorrelates the fast turbulent internal dynamics of stripes from slow processes such as their proliferation and decay \cite{avila2011onset,shi2013scale,gome2020statistical}. Moreover, the Hurst exponent of 1/2 marks such slow processes as stochastic random events, a key requirement for the statistical nature of the percolation phase transition \cite{lemoult2016directed,chantry2017universal,klotz2022phase,hof2023directed} encountered at higher \Rey{} in many shear flows.

Hydrodynamic stability concepts developed more than a century ago allowed for the identification of the first bifurcation to a non-trivial vortex state \cite{taylor1923stability}, and with it the starting point for the supercritical route to turbulence in linearly unstable flows. Finding corresponding flow states for the much more volatile transition characteristic of most flows of practical relevance, such as pipe and channel flows, has proven far more difficult. Exploiting that the statistics of turbulence are generic and independent of the numerical domain at sufficiently high \Rey{} (see SM \cite{Note1}), we selected a domain that stabilizes stripes in the transitional regime of channel flow. The stripe solutions identified in this configuration are spatially periodic in the stripe direction and hence differ from the doubly localized stripes observed in experiments close to the critical point. It is likely that doubly localized stripe solutions \cite{zammert2014streamwise,kanazawa2018thesis} bifurcate from the ECS presented in this study. However, the continuous route from turbulence to ECSs identified here can only be established by bypassing the doubly localized stripe regime, in which flows unavoidably relaminarize (as illustrated in \reffig{cartoon}). Although the tilted domain may appear specific, the two states shown to be dynamically connected, i.e.\ the periodic orbit and fully turbulent flow, are generic to the classic channel flow problem and entirely independent of this particular choice.

A striking feature of the route towards turbulence is the abruptness of the dimension change directly at the onset of chaos, long before turbulence is observable in experiments. This steep dimension increase marks the border up to which deterministic concepts are suitable whereas above statistical mechanics descriptions
become more appropriate, setting the stage for the non-equilibrium phase transition \cite{avila2011onset,lemoult2016directed} encountered at larger \Rey.

\begin{acknowledgments}
We thank Baofang Song as well as the developers of Channelflow for sharing their numerical codes, and Mukund Vasudevan and Holger Kantz for fruitful discussions. This work was supported by a grant from the Simons Foundation (662960, BH).
\end{acknowledgments}

\bibliographystyle{apsrev4-2}
\bibliography{ms}

\onecolumngrid
\clearpage
\widetext

\begin{center}
	\textbf{\large Supplemental Material}
\end{center}
\setcounter{equation}{0}
\setcounter{figure}{0}
\setcounter{table}{0}

\renewcommand{\theequation}{S\arabic{equation}}
\renewcommand{\thefigure}{S\arabic{figure}}
\renewcommand{\thetable}{S\Roman{table}}

\section{Coordinates}
\begin{figure}
\begin{overpic}[height=0.45\linewidth]{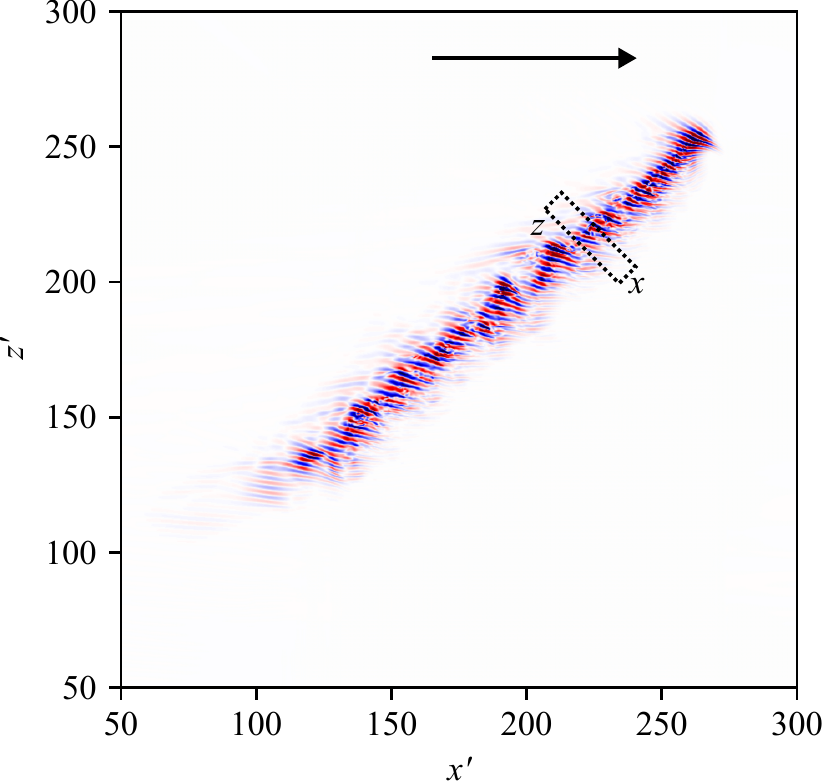}
  \put (-5, 95) {(a)}
\end{overpic}
\begin{overpic}[height=0.45\linewidth]{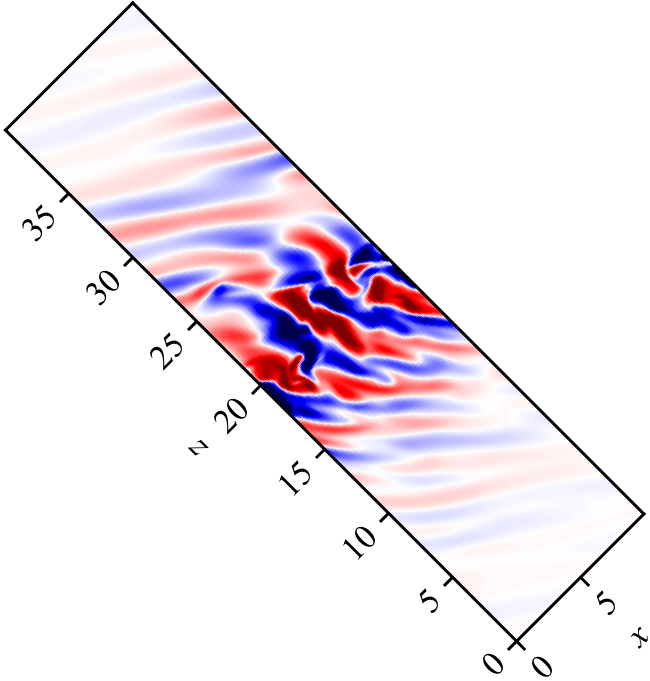}
  \put (0, 98) {(b)}
\end{overpic}
\caption{Illustration of the tilted domain. Colors show the wall-normal velocity ($v$) contours (range limited from $v=-0.1$ in blue to $v=0.1$ in red) in the midplane. (a) Large non-tilted periodic domain (sized $400\times 400$, here zoomed-in to an area of $250\times250$), with an isolated turbulent stripe $\Rey{}=660$. The dashed box, tilted by $\theta=45^\circ$ with respect to the streamwise direction, describes the $10\times40$ sized periodic tilted domain used in our simulations and its axes $(x, z)$. The arrow shows the streamwise direction $\bm{e}_{x'}$. (b) Narrow periodic tilted domain (sized $10\times40$) used in our simulations, with a turbulent stripe at $\Rey{}=660$.}
\label{fig:tilted_concept}
\end{figure}

In order to capture the long time dynamics of a turbulent stripe at a relatively lower cost, we used the tilted domain trick first used in plane Couette flow and later adapted to plane Poiseuille flow \cite{tuckerman2020patterns}. \reffig{tilted_concept} illustrates this tilted domain: the small tilted box drawn on the stripe within the large periodic domain shows the configuration of the tilted axes $(x, z)$, by convention $L_z>L_x$. We non-dimensionalize space with the half-gap length $h$, and center the wall-normal coordinate $y$ on the midplane, therefore $y\in[-1,1]$. The tilted box is rotated counter-clockwise by $\theta$ with respect to streamwise direction such that its short side (the $x$ axis) becomes parallel to the turbulent stripe. The relation between the unit vectors of the two coordinate systems, $(x',z')$, parallel to the streamwise and spanwise directions respectively, and $(x, z)$, parallel to the short and long sides of the tilted periodic domain respectively, is then
\begin{equation}
\begin{aligned}
  \bm{e}_{x'}& = \cos \theta \bm{e}_x - \sin \theta \bm{e}_z\,,\\
  \bm{e}_{z'}& = \sin \theta \bm{e}_x + \cos \theta \bm{e}_z\,.
\end{aligned}
\label{eq:directions}
\end{equation}

In our simulations we time evolve the perturbation velocity $\bm{u}$ which is the difference of the total velocity field $\bm{U}$ from the laminar solution. We use the value of streamwise component of the laminar solution at the midplane, $\bm{U}_\mathrm{laminar}(0)\cdot\bm{e}_{x'}$, as the velocity scale. Therefore $\bm{u} = \bm{U} - (1-y^2)\bm{e}_{x'}$. We refer to the projections of perturbation velocity onto the directions defined in \refeq{directions} with primed letters for the streamwise perturbation velocity $u'=\bm{u} \cdot \bm{e}_{x'}$ and spanwise perturbation velocity $w'=\bm{u} \cdot \bm{e}_{z'}$, and with non-primed letters for the directions of the simulation domain  $u=\bm{u} \cdot \bm{e}_{x}$ and $w=\bm{u} \cdot \bm{e}_{z}$. The wall-normal component is the same in both cases, $v'=v=\bm{u} \cdot \bm{e}_{y}$.

Later we present visualizations of streamwise vorticity,
\begin{equation}
  \omega' = (\nabla'\times\bm{u})\cdot \bm{e}_{x'} = \frac{\partial w'}{\partial y'} - \frac{\partial v'}{\partial z'}\,,
\end{equation}
for which one needs to know how the derivative $\partial/\partial_{z'}$ transforms:
\begin{equation}
  \frac{\partial}{\partial z'} = \sin \theta \frac{\partial}{\partial x} + \cos \theta \frac{\partial}{\partial z}\,.
\end{equation}

\section{Invariant solutions}
\begin{figure}
  \begin{overpic}[width=0.35\linewidth]{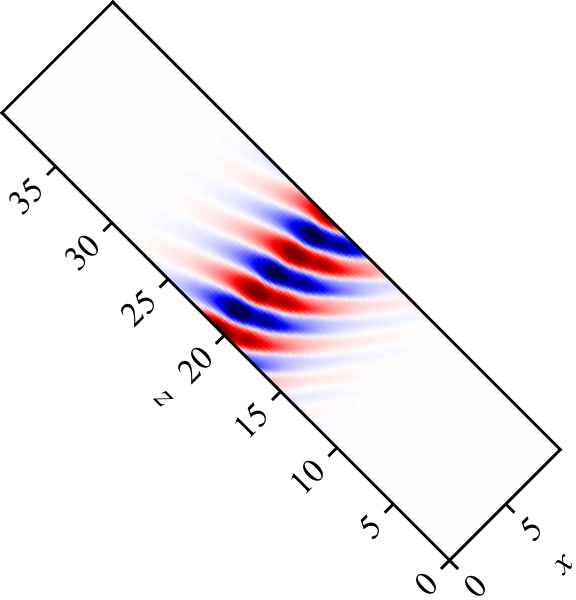}
    \put (0, 100) {(a)}
  \end{overpic}
  \begin{overpic}[width=0.64\linewidth]{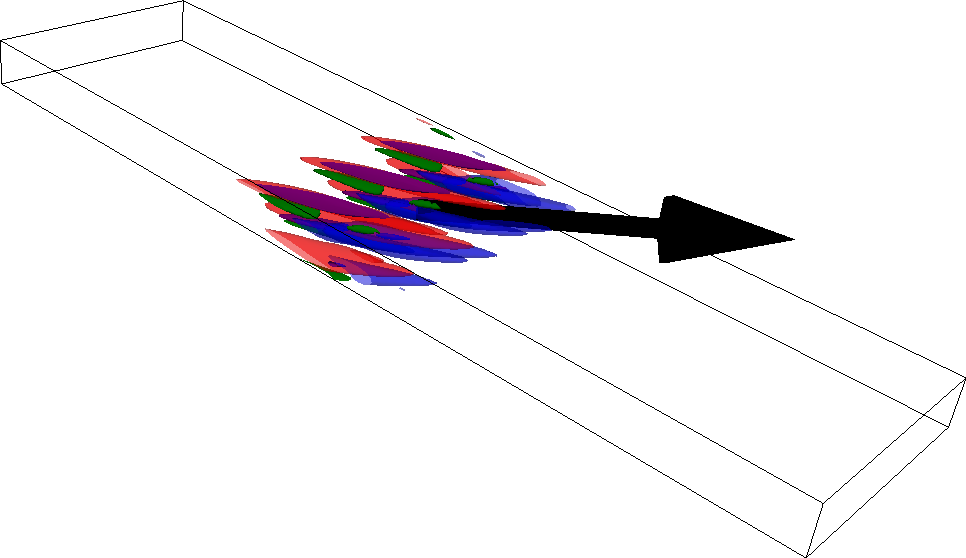}
    \put (0, 56.5) {(b)}
  \end{overpic}\\
  \begin{overpic}[width=0.35\linewidth]{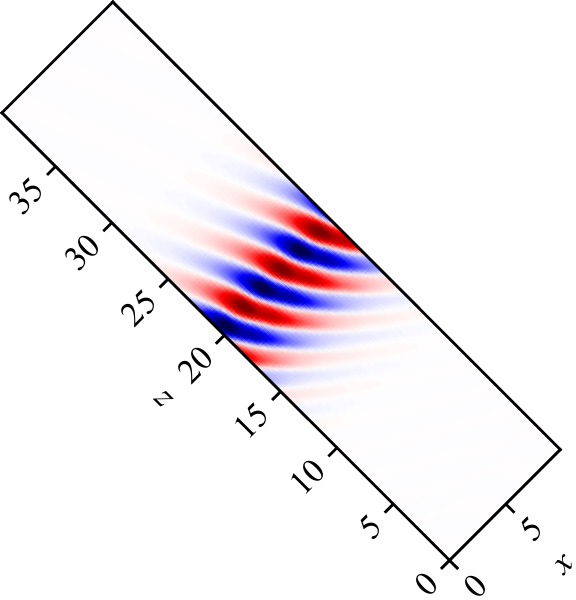}
    \put (0, 100) {(c)}
  \end{overpic}
  \begin{overpic}[width=0.64\linewidth]{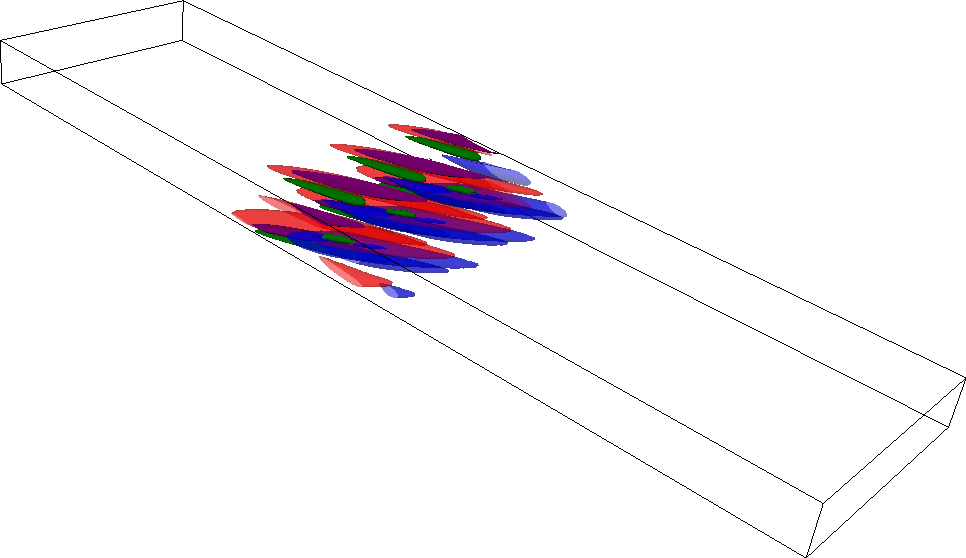}
    \put (0, 56.5) {(d)}
  \end{overpic}\\
  \begin{overpic}[width=0.35\linewidth]{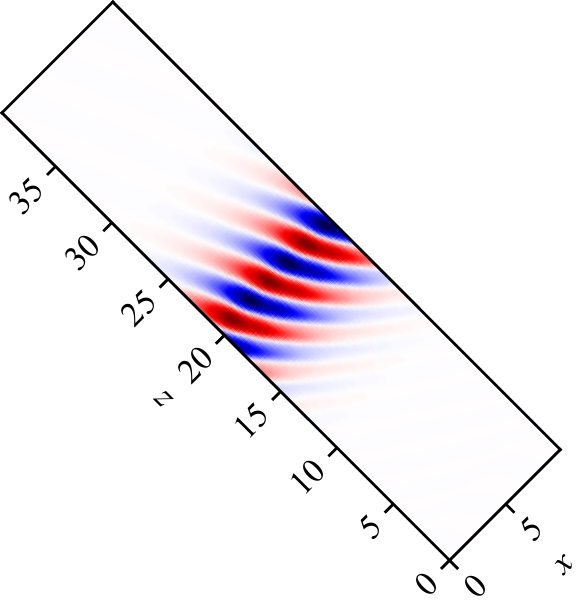}
    \put (0, 100) {(d)}
  \end{overpic}
  \begin{overpic}[width=0.64\linewidth]{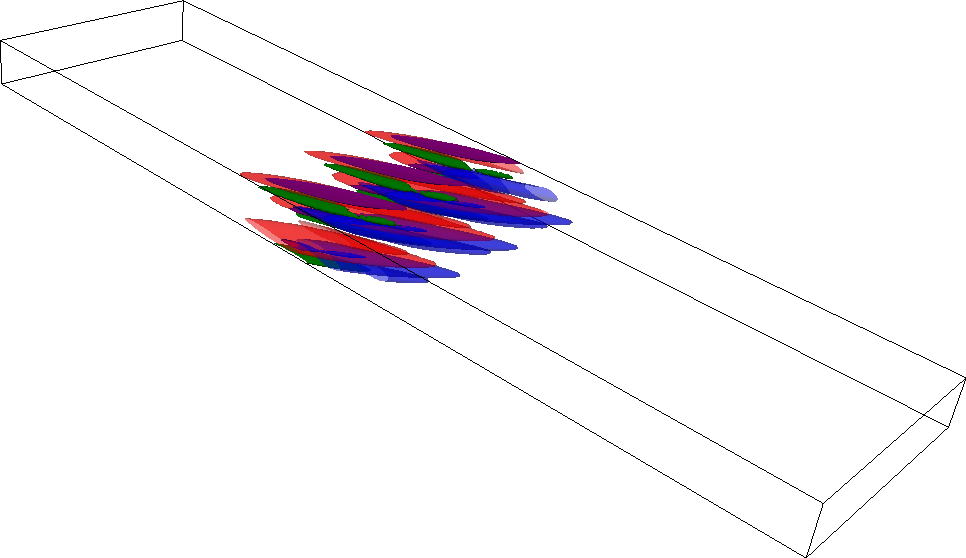}
    \put (0, 56.5) {(e)}
  \end{overpic}
  \caption{Traveling waves at \Rey{} (a,b) 371, (c,d) 380 and (e,f) 387. Left column: Contours of wall-normal velocity ($v$) at the midplane, range limited from $v=-0.1$ in blue to $v=0.1$ in red. Right column: Isosurfaces of $0.5\min / \max u'$ (blue/red) and $0.5\min / \max \omega'$ (purple/green). The arrow shows the streamwise direction $\bm{e}_{x'}$.}
  \label{fig:TWs}
\end{figure}

\begin{figure}
  \begin{overpic}[width=0.35\linewidth]{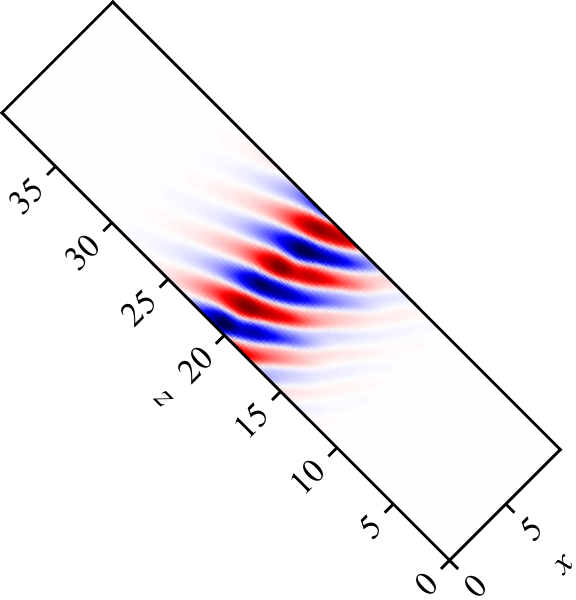}
    \put (0, 100) {(a)}
  \end{overpic}
  \begin{overpic}[width=0.64\linewidth]{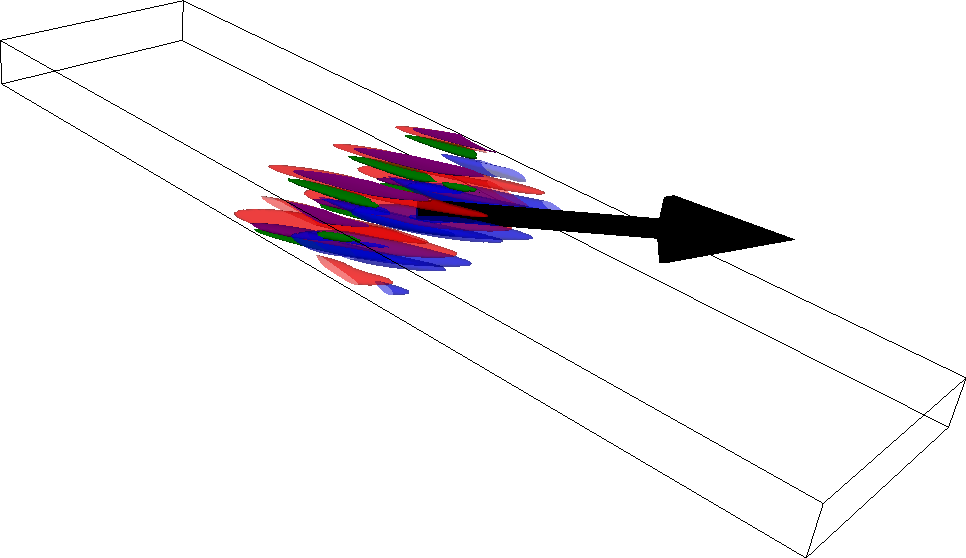}
    \put (0, 56.5) {(b)}
  \end{overpic}\\
  \begin{overpic}[width=0.35\linewidth]{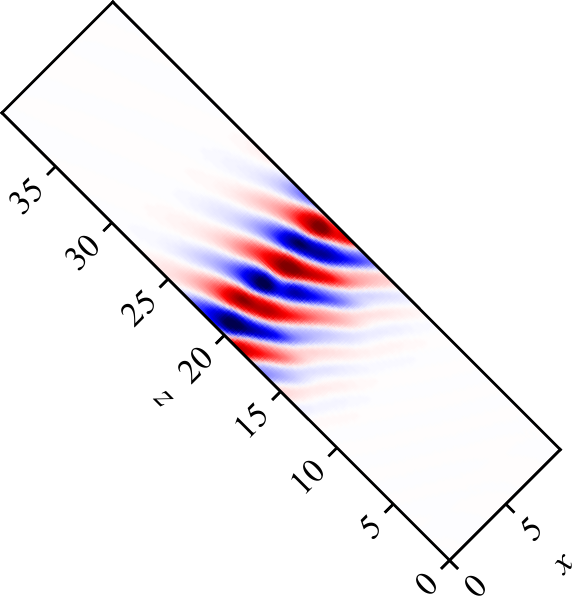}
    \put (0, 100) {(c)}
  \end{overpic}
  \begin{overpic}[width=0.64\linewidth]{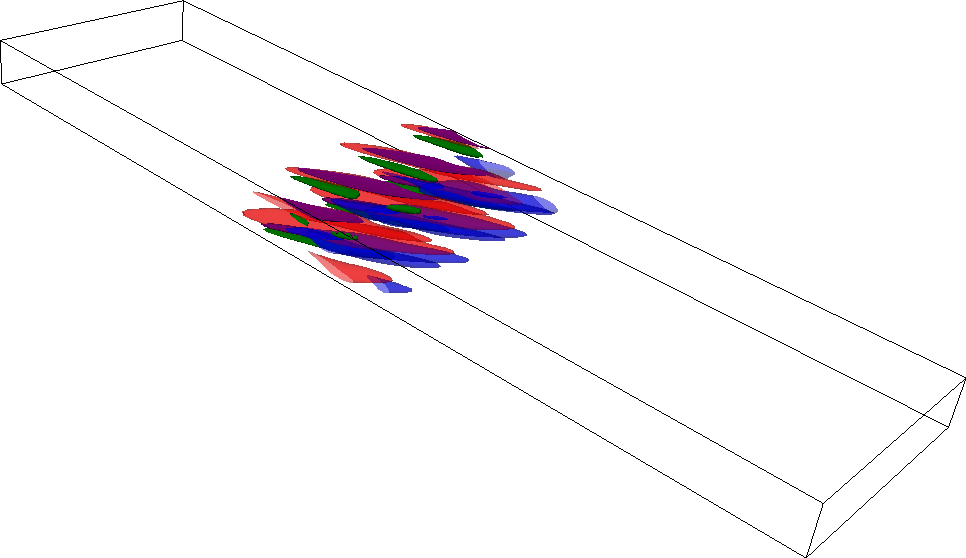}
    \put (0, 56.5) {(d)}
  \end{overpic}\\
  \begin{overpic}[width=0.35\linewidth]{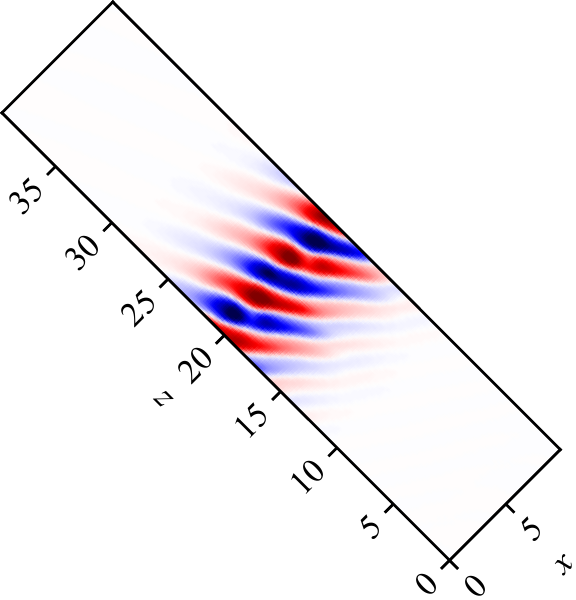}
    \put (0, 100) {(e)}
  \end{overpic}
  \begin{overpic}[width=0.64\linewidth]{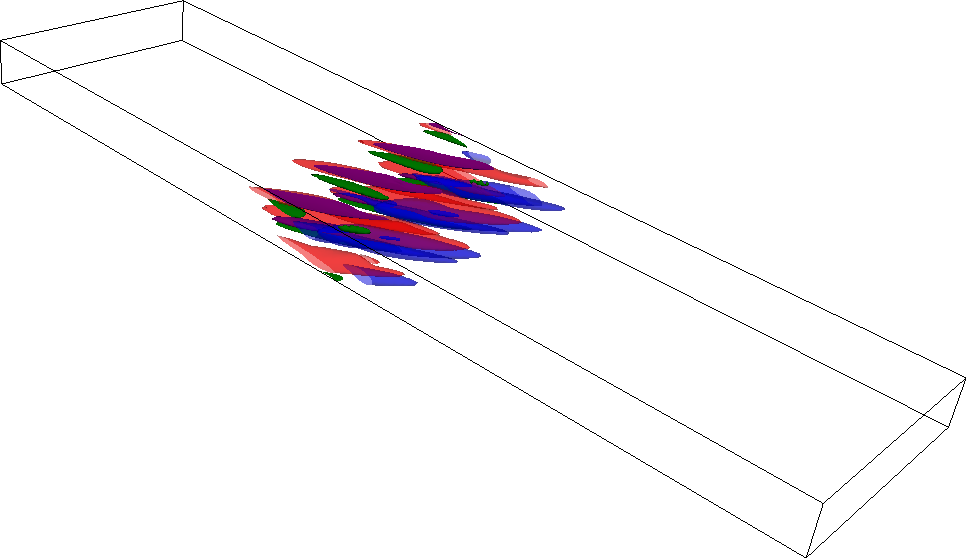}
    \put (0, 56.5) {(f)}
  \end{overpic}
  \caption{Unstable periodic orbits at \Rey{} (a,b) 388.2, (c,d) 390.1, (e,f) 391.8. Left column: Contours of wall-normal velocity ($v$) at the midplane, range limited from $v=-0.1$ in blue to $v=0.1$ in red. Right column: Isosurfaces of $0.5\min / \max u'$ (blue/red) and $0.5\min / \max \omega'$ (purple/green). The arrow shows the streamwise direction $\bm{e}_{x'}$.}
  \label{fig:uRPOs}
\end{figure}

Here we provide some visualizations of the invariant solutions we found in the $10\times40$ domain, the spectral discretization of which assumes $96 \times 256$ Fourier modes in the plane and 49 collocation points in the wall-normal direction. The solutions were converged with the implementation of the Newton--Krylov algorithm \cite{viswanath2007recurrent} in Channelflow \cite{channelflow2}. \reffig{TWs} shows traveling waves: solutions $\bm{u}$ that obey $\bm{u}(t,\, x,\, y,\, z) = \bm{u}(t+T,\, x - c_x T,\, y,\, z - c_z T)$ for all $T$ and specific phase velocities $c_x$ and $c_z$.
\reffigs{uRPOs}{sRPOs} show (unstable and stable, respectively) periodic orbits: solutions $\bm{u}$ that obey $\bm{u}(t,\, x,\, y,\, z) = \bm{u}(t + T,\, x - \sigma_x,\, y,\, z - \sigma_z)$ for specific $T$ and shifts $\sigma_x$ and $\sigma_z$. $T$ is called the period when it is the minimum positive value that fulfills this relation, any integer multiple of the period with corresponding shifts also fulfills this relation. We computed the stability of these solutions using the Arnoldi iteration implemented in Channelflow \cite{channelflow2}.

The periodic orbits found in the \Rey{} descent are `relative' periodic orbits: their shifts $\sigma_x$ and $\sigma_z$ as defined above are nonzero. They are stable, albeit in a narrow \Rey{} window only, and develop unstable directions at lower \Rey{} ($\ReyuRPO=393$, see \reftab{reynolds_full} for a list of transition points encountered for decreasing \Rey{}, including the regime down \ReyuRPO{}). Despite turning unstable, the periodic orbit can still be tracked further using a Newton--Krylov algorithm \cite{viswanath2007recurrent,channelflow2} and is found to bifurcate from a traveling wave (TW) at $\ReyTW=387.6$. 
\begin{table}
	\caption{Named Reynolds numbers and observations below the given \Rey{} in simulations in the $10 \times 40$ sized tilted domain, an extended version of \reftab{reynolds}. Subscripts refer to the observations below the given \Rey{} whereas superscripts refer to the observations above.}
    \label{tab:reynolds_full}
	\begin{ruledtabular}
		\begin{tabular}{l r l}
            Name & Value & Observations below \Rey{} \\
            \hline
            \Reypatch{}      & $\approx 1500$   & Spatio-temporal intermittency\\[1mm]
            \Reytransient{}  & $\approx 650$    & Transient chaos\\[1mm]
            \Reyattractor{}  & $\approx 420$    & Sustained chaos\\[1mm]
            \Reytorus{}      & $412.8$          & Quasiperiodicity\\[1mm]
            \ReysRPO{}       & $403.2$          & Stable periodic orbits\\[1mm]
            \ReyuRPO{}       & $393$            & Unstable periodic orbits\\[1mm]
            \ReyTW{}         & $387.6$          & Traveling waves\\[1mm]
            \Reysaddlenode{} & $370.6$          & Laminar flow
		\end{tabular}
	\end{ruledtabular}
\end{table}
This unstable traveling wave, an edge state previously identified in \cite{paranjape2019thesis, paranjape2020oblique}, again has the form of a localized stripe. The TW can equally be tracked to lower \Rey{} and it finally disappears in a saddle-node bifurcation ($\Reysaddlenode=370.6$), below which we could not identify any invariant solution and observed laminar Poiseuille flow only. Unlike most studies restricted to symmetry subspaces \cite{kreilos2012periodic,avila2013streamwise}, in the present case it is not the upper but the lower branch TW that gives rise to chaos.

\begin{figure}
  \begin{overpic}[width=0.35\linewidth]{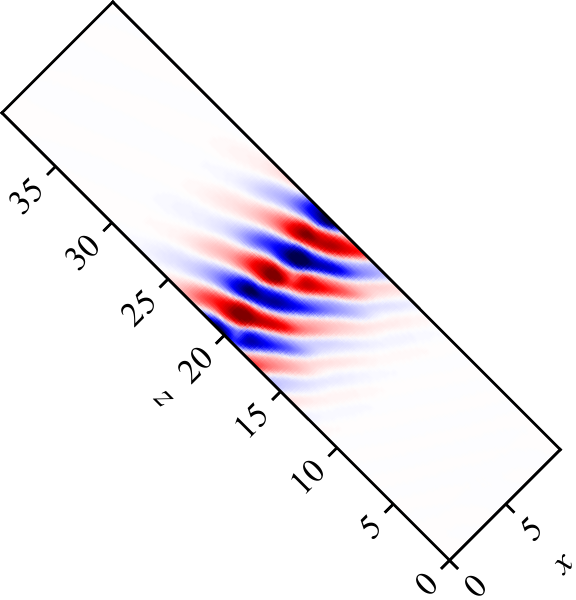}
    \put (0, 100) {(a)}
  \end{overpic}
  \begin{overpic}[width=0.64\linewidth]{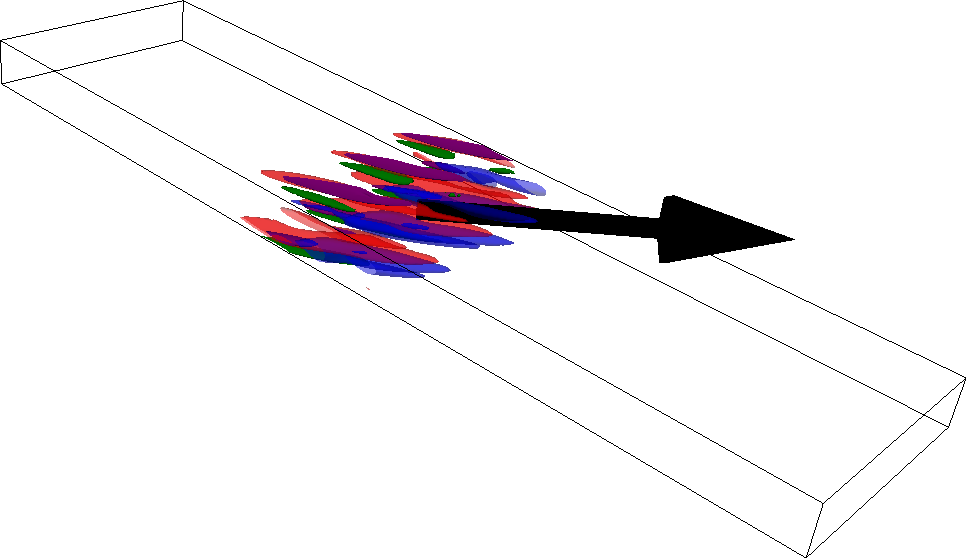}
    \put (0, 56.5) {(b)}
  \end{overpic}\\
  \begin{overpic}[width=0.35\linewidth]{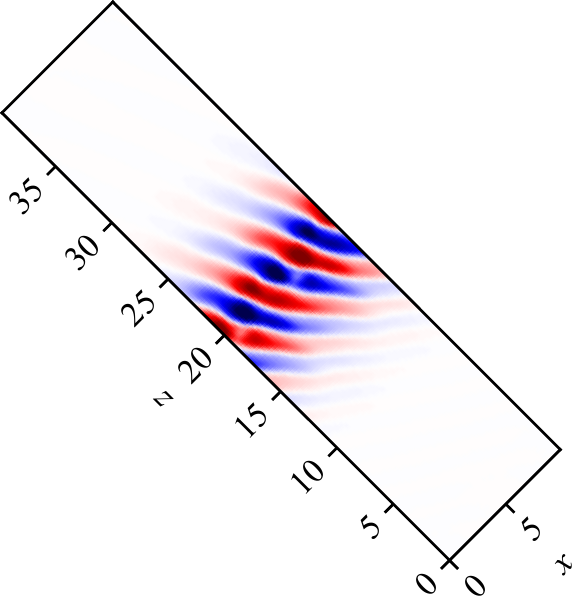}
    \put (0, 100) {(c)}
  \end{overpic}
  \begin{overpic}[width=0.64\linewidth]{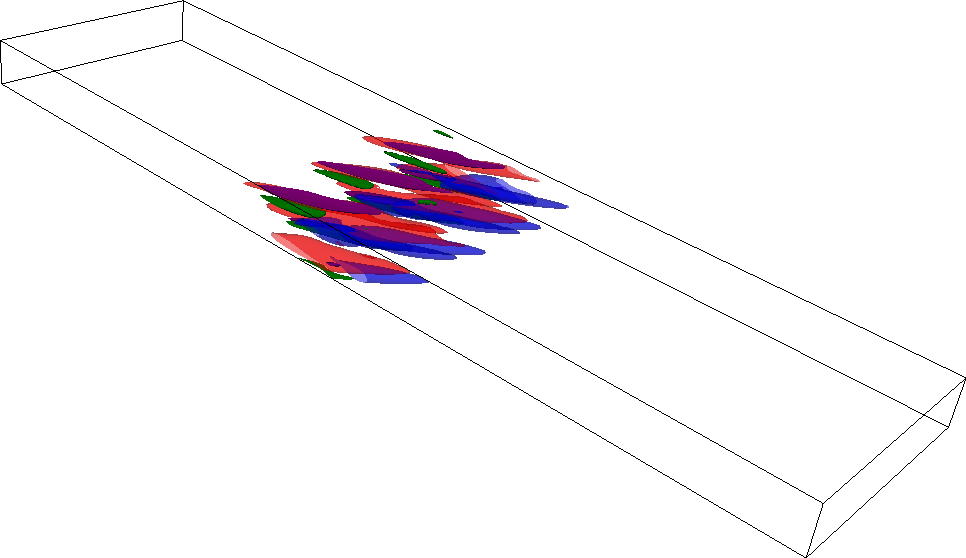}
    \put (0, 56.5) {(d)}
  \end{overpic}\\
  \begin{overpic}[width=0.35\linewidth]{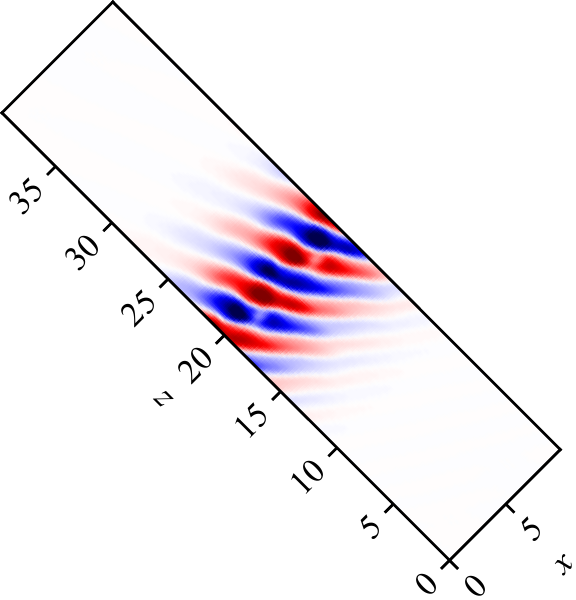}
    \put (0, 100) {(e)}
  \end{overpic}
  \begin{overpic}[width=0.64\linewidth]{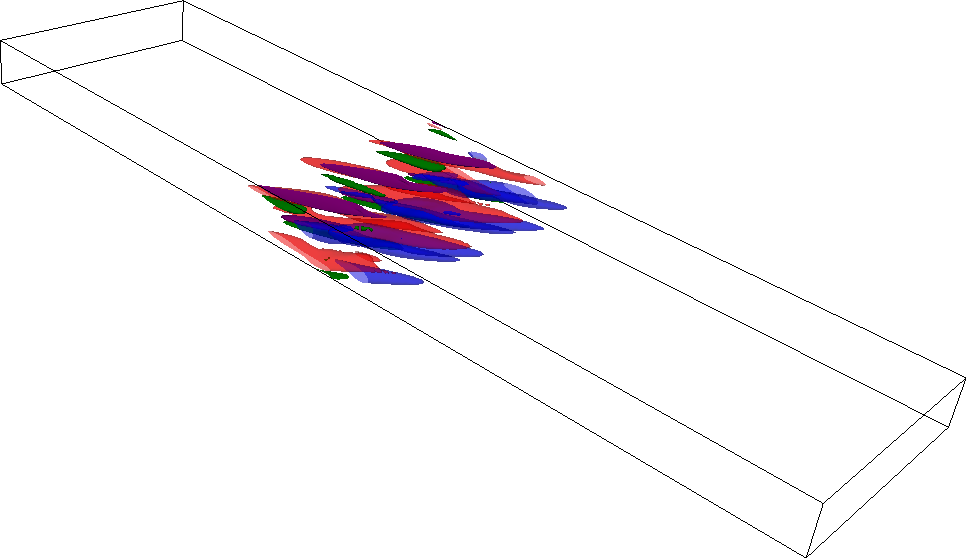}
    \put (0, 56.5) {(f)}
  \end{overpic}
  \caption{Stable periodic orbits at \Rey{} (a,b) 395, (c,d) 398.6, (e,f) 401.7. Left column: Contours of wall-normal velocity ($v$) at the midplane, range limited from $v=-0.1$ in blue to $v=0.1$ in red. Right column: Isosurfaces of $0.5\min / \max u'$ (blue/red) and $0.5\min / \max \omega'$ (purple/green). The arrow shows the streamwise direction $\bm{e}_{x'}$.}
  \label{fig:sRPOs}
\end{figure}

\section{The case of $L_z=120$}
In order to confirm that the bifurcation sequence we observed (\reffig{bifurcations}) is not specific to the domain size we chose, we tripled the domain size in $L_z$ and studied the bifurcation sequence observed in a periodic domain of size $10\times 120$. The spectral discretization of this larger domain assumes $96 \times 768$ Fourier modes in the plane and 49 collocation points in the wall-normal direction. 

\begin{figure}
  \includegraphics[width=0.55\linewidth]{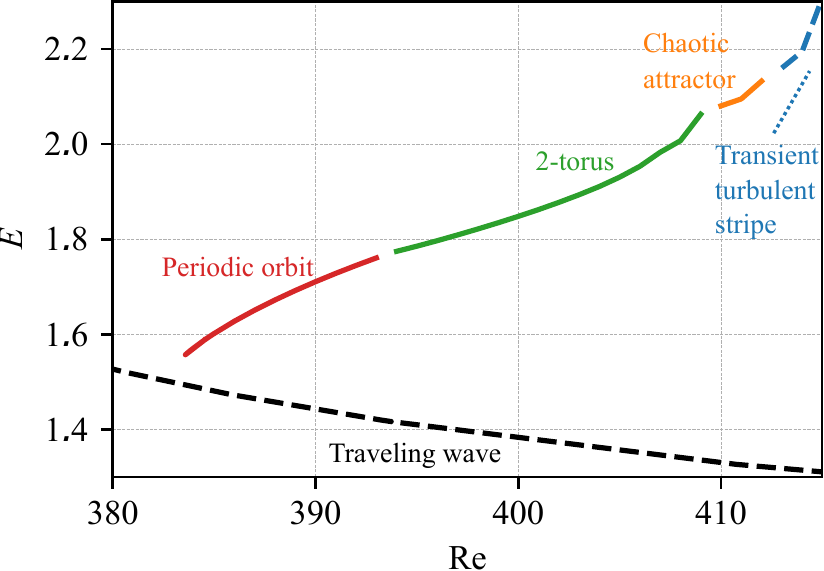}
  \caption{Time-averaged perturbation kinetic energy $E$ vs.\ \Rey{} during the descent in the $10\times120$ domain. Annotations refer to the respective dynamics encountered as \Rey{} is decreased. Larger-domain counterpart of \reffig{bifurcations}(a).}
  \label{fig:bifurcations_Lz120}
\end{figure}

The analysis of the correlation dimension $D_2$ (\refeq{D2}) is costly as we discussed in the Letter.
In this three times larger domain we did not calculate it. However, we observed the same qualitative sequence seen in the $10\times40$ domain. As we decreased \Rey{}, starting from a turbulent stripe at $\Rey{}=700$, we saw: transient chaos, sustained chaos, tori, and stable periodic orbits. See \reffig{bifurcations_Lz120} for a plot of the perturbation kinetic energy during this descent and \reftab{reynolds_lz120} for a list of the transition points. We note that while the specific values of \Rey{} where transitions happen are different between the two domains, the transitions and their order are the same. While we did continue the stable periodic orbits further towards lower \Rey{} and found unstable periodic orbits, we did not wait for the continuation to get to its limit. The continuation process gets slower, and therefore costlier, as \Rey{} gets nearer to a bifurcation point. We expect from \reffig{bifurcations_Lz120} that the unstable periodic orbits should again bifurcate from a traveling wave at lower \Rey{}. In order to identify the traveling waves, we instead ran a bisection algorithm at higher \Rey{} and continued the thereby found traveling wave down in \Rey{} with arclength continuation. The bisection algorithm is also implemented in Channelflow \cite{channelflow2}.

\begin{table}
	\caption{Named Reynolds numbers and observations below the given \Rey{} in simulations in the $10 \times 120$ sized tilted domain. Subscripts refer to the observations below the given \Rey{} whereas superscripts refer to the observations above. Larger domain counterpart of \reftab{reynolds_full}.}
    \label{tab:reynolds_lz120}
	\begin{ruledtabular}
		\begin{tabular}{l r l}
            Name & Value & Observations below \Rey{} \\
            \hline
            \Reyattractor{}  & $\approx 413$            & Sustained chaos\\[1mm]
            \Reytorus{}      & $408.8$                  & Quasiperiodicity\\[1mm]
            \ReysRPO{}       & $393$                    & Stable periodic orbits\\[1mm]
            \ReyuRPO{}       & $387$                    & Unstable periodic orbits\\[1mm]
            \ReyTW{}         & $\approx 383.5$          & Travelling waves
		\end{tabular}
	\end{ruledtabular}
\end{table}

\section{Simulations of a fully turbulent flow, $\Rey{}=4200$}
We simulated turbulence at values of \Rey{} higher than where sustained turbulent stripes are found in order to have a complete path in \Rey{} between invariant solutions at low \Rey{} and fully turbulent flow at higher \Rey{}. In particular, we started at $\Rey=4200$, following the fully-resolved simulations of \refref{kim1987turbulence} (at $\Rey_\tau=180$, see \refeq{Retau}) in a non-tilted domain of size $4\pi \times 2\pi$ ($L_x' \times L_z'$) which was discretized with $192\times129$ Fourier modes in the plane and $129$ collocation points in the wall-normal direction. 

In order to have an illustrative visualization (\reffig{cartoon}) with an isolated turbulent stripe in addition to a stripe in our tilted domain, we ran a non-tilted $\Rey=4200$ simulation in a square domain of size $50 \cos 45^\circ \times 50 \cos 45^\circ \approx 35.4 \times 35.4$ ($L_x'\times L_z'$), the rectangle of minimal size that fits a 45 degree tilted $10\times40$ rectangle. We discretized this domain with $540\times 900$ Fourier modes in the plane and $135$ collocation points in the wall-normal direction.

\begin{figure}
  \begin{overpic}[width=0.45\linewidth]{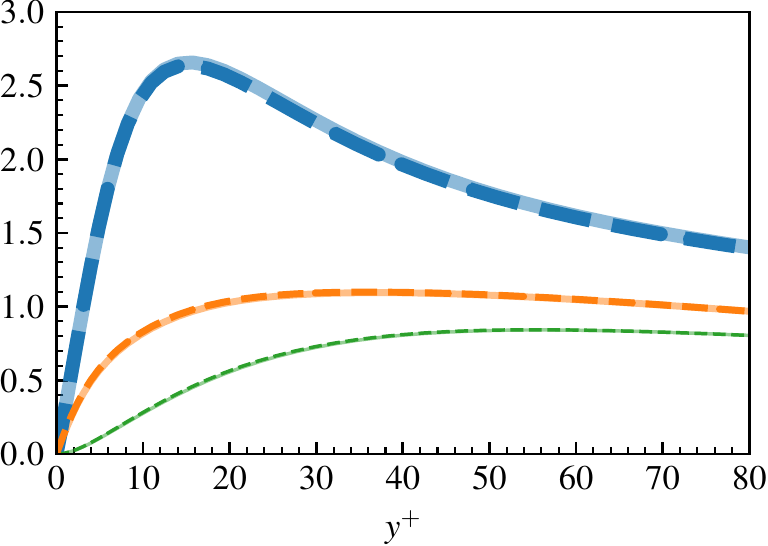}
    \put (-15, 42) {\color{blue}{$u_\mathrm{rms}/u_\tau$}}
    \put (-15, 37) {\color{orange}{$w_\mathrm{rms}/u_\tau$}}
    \put (-15, 32) {\color{green}{$v_\mathrm{rms}/u_\tau$}}
  \end{overpic}
  \caption{Time-averaged root-mean-squared velocity fluctuations (\refeq{rmsvels}) at $\Rey=4200$ normalized with wall-shear velocity ($u_\tau$, \refeq{frictionvel}) as a function of the distance to the wall in wall units ($\delta_\nu$): top/blue line is for streamwise, middle/orange is for spanwise, and bottom/green line is for wall-normal velocity fluctuations. Solid transparent lines are for the $10\times40$ sized tilted domain, dashed lines are for the $35.4 \times 35.4$ sized non-tilted domain. Figure 6(b) of \refref{kim1987turbulence} plots the same quantities at the same $\Rey_\tau=180$ (\refeq{Retau}) in a smaller, $4\pi \times 2\pi$ sized non-tilted domain.}
  \label{fig:rms_vels}
\end{figure}

As a basic check that turbulence in the small tilted box ($10\times40$, at $\Rey=4200$, discretized with $180\times 720$ Fourier modes in the plane and $135$ collocation points in the wall-normal direction) is comparable to turbulence in the larger, non-tilted box ($35.4\times35.4$), we computed velocity fluctuations around the mean,
\begin{equation}
  \begin{aligned}
    u_\mathrm{rms}^2(y) &= \frac{1}{t_f-t_i} \int_{t_i}^{t_f} \mathrm{d}t\, \frac{1}{L_x L_z} \int_0^{L_x} \int_0^{L_z} \mathrm{d}x\,\mathrm{d}z\, \{[\bm{u}(t,x,y,z)-\langle \bm{u} \rangle(y)]\cdot \bm{e}_{x'}\}^2\,,\\
    v_\mathrm{rms}^2(y) &= \frac{1}{t_f-t_i} \int_{t_i}^{t_f} \mathrm{d}t\, \frac{1}{L_x L_z} \int_0^{L_x} \int_0^{L_z} \mathrm{d}x\,\mathrm{d}z\, \{[\bm{u}(t,x,y,z)-\langle \bm{u} \rangle(y)]\cdot \bm{e}_{y}\}^2\,,\\
    w_\mathrm{rms}^2(y) &= \frac{1}{t_f-t_i} \int_{t_i}^{t_f} \mathrm{d}t\, \frac{1}{L_x L_z} \int_0^{L_x} \int_0^{L_z} \mathrm{d}x\,\mathrm{d}z\, \{[\bm{u}(t,x,y,z)-\langle \bm{u} \rangle(y)]\cdot \bm{e}_{z'}\}^2\,,
  \end{aligned}
  \label{eq:rmsvels}
\end{equation}
and plotted these fluctuations as a function of the distance to the wall in units of the viscous length (\refeq{lviscous}), see \reffig{rms_vels}. Note that as an exception we withheld the primes above the streamwise/spanwise components $u'$/$w'$ for clarity. The resulting curves of the two domains agree with each other, as well as with the corresponding results of \refref{kim1987turbulence}, Figure 6(b) therein.

Computing these velocity fluctuations requires one to first calculate the mean velocity,
\begin{equation}
  \langle \bm{u} \rangle(y) = \frac{1}{(t_f-t_i)} \int_{t_i}^{t_f} \mathrm{d}t\, \frac{1}{L_x L_z} \int_0^{L_x} \int_0^{L_z} \mathrm{d}x\,\mathrm{d}z\, \bm{u}(t,x,y,z)\,,
\end{equation}
for which we took a $\approx 13 h / u_\tau$ long ($u_\tau$ is the friction velocity, \refeq{frictionvel}) trajectory each in both the non-tilted domain and the tilted domain. From the mean velocity, one can compute the wall-shear stress,
\begin{equation}
  \tau_w = \rho \nu \left.\frac{\mathrm{d}\langle \bm{U} \rangle \cdot \bm{e}_{x'}}{\mathrm{d}y}\right|_{y=-h}\,,
\end{equation}
where $\bm{U}$ is the total velocity. Written in terms of the perturbation velocity $\bm{u}$ and non-dimensionalized in terms of the half-gap $h$, kinematic viscosity $\nu$, density $\rho$ and laminar centerline velocity, this equation gives
\begin{equation}
  \tau_w^\star = \frac{2}{\Rey{}} + \frac{1}{\Rey{}}\left.\frac{\mathrm{d}\langle \bm{u} \rangle \cdot \bm{e}_{x'}}{\mathrm{d}y}\right|_{y=-1}\,,
\end{equation}
where $\star$ denotes that the variable is non-dimensionalized.
Using the wall-shear stress $\tau_w$, one can further define a new velocity scale, called the friction velocity,
\begin{equation}
    u_\tau = \sqrt{\frac{\tau_w}{\rho}}\,,
\label{eq:frictionvel}
\end{equation}
which in non-dimensional form is
\begin{equation}
  u_\tau^\star = \sqrt{\tau_w^\star}\,.
\end{equation}
Additionally, one can define a new length scale, called the viscous length,
\begin{equation}
    \delta_v = \nu \sqrt{\frac{\rho}{\tau_w}} = \frac{\nu}{u_\tau}\,,
\label{eq:lviscous}
\end{equation}
which in non-dimensional form is
\begin{equation}
  \delta_\nu^\star = 1/(u_\tau^\star \Rey{})\,.
\end{equation}
This defines the ``inner units'': $y^+$ in \reffig{rms_vels} is the distance to the wall in units of the viscous length $\delta_\nu$. 

The friction velocity can be used to define the friction Reynolds number,
\begin{equation}
    \Rey_\tau = \frac{u_\tau h}{\nu} = \frac{h}{\delta_\nu}\,,
\label{eq:Retau}
\end{equation}
which can also be given in terms of the non-dimensionalized friction velocity and the ``outer units'' Reynolds number \Rey{} used everywhere else in this work,
\begin{equation}
  \Rey_\tau = u_\tau^\star \Rey{}\,.
\end{equation}
Note that as $u_\tau^\star$ itself is dependent on \Rey{}, this equation does not imply a proportionality between $\Rey_\tau$ and \Rey{}.

\end{document}